\documentclass[onecolumn,showpacs]{article}
\usepackage{adjustbox}
\usepackage{graphicx}
\usepackage{dcolumn}
\usepackage{bm}
\usepackage{amsmath}
\usepackage{subcaption}
\usepackage{amssymb}
\usepackage{cite}
\usepackage{tablefootnote}
\usepackage{multirow}
\usepackage{caption}
\usepackage{epstopdf}
\usepackage{hyperref}
\usepackage{wrapfig}
\usepackage{lipsum} 

\begin{document}
{\setlength{\oddsidemargin}{1.2in}
\setlength{\evensidemargin}{1.2in} } \baselineskip 0.55cm
\begin{center}
{\LARGE {\bf Evolutionary Behavior of Fractional Holographic Dark Energy within $f(T)$ Teleparallel Gravity}}
\end{center}
\date{\today}
\begin{center}
  Elangbam Chingkheinganba Meetei$^1$, S. Surendra Singh$^2$ \\
   $^{1,2}$Department of Mathematics, National Institute of Technology Manipur, Imphal, India\\
   Email: $^1$chingelang@gmail.com, $^2$ssuren.mu.ac.in
 \end{center}

 \textbf{Abstract:} We investigate the cosmological dynamics of FHDE within $f(T)$ gravity by employing the dynamical system approach in a spatially flat FRW background. By introducing appropriate dimensionless variables, the field equations are reformulated as a closed system, which allows a systematic phase-space analysis. The resulting system admits four critical points, including two saddle points corresponding to radiation and matter-dominated epochs, and two stable points associated with a DE-dominated phase and a de Sitter solution. The radiation- and matter-dominated critical points are found to possess a saddle character in phase space, ensuring their transient nature and enabling the cosmological evolution to naturally progress toward a stable late-time accelerated attractor. The stable critical points describe accelerated expansion with effective equations of state compatible with DE and de Sitter regimes. Overall, the analysis indicates that $f(T)$ gravity is capable of reproducing the standard cosmological sequence within a consistent dynamical framework.
 
 
 \section{Introduction}
The discovery of the late-time accelerated expansion of the Universe, confirmed through observations of Type Ia supernovae, cosmic microwave background anisotropies, and baryon acoustic oscillations, has led to intense research on the nature of dark energy (DE) \cite{Riess1998,Perlmutter1999,Planck2016}. Although the $\Lambda$CDM model, consisting of a cosmological constant $\Lambda$ and cold dark matter, provides an excellent fit to most observational datasets, it suffers from several unresolved theoretical issues-including the fine-tuning and cosmic coincidence problems \cite{Weinberg1989,Sahni2000}. These shortcomings have motivated the cosmology community to explore dynamical alternatives to $\Lambda$, such as scalar-field models, holographic dark energy (HDE), and various modified gravity theories.

The holographic principle, originally proposed within the framework of black hole thermodynamics and quantum gravity, suggests that the number of degrees of freedom scales with the boundary area rather than the volume \cite{Bekenstein1973,tHooft1993,Susskind1995,Cohen1999}. Based on this principle, Li formulated the HDE model where the energy density of DE is inversely related to the square of an infrared (IR) cutoff length scale $L$ \cite{Li2004}. However, the standard HDE model still faces challenges-particularly in explaining the transition from deceleration to acceleration for certain IR cutoffs.

Motivated by developments in fractional calculus, a generalized extension known as fractional holographic dark energy (FHDE) has been introduced. Here, the fractional-order modifications alter the dependence of $\rho_{FHDE}$ on the IR cutoff, thereby enriching the dynamical behavior of the model and potentially addressing some of the limitations of standard HDE \cite{Bamba2014,Nunes2016,Tavayef2018,Vagenas2019}. Depending on the fractional parameter and the choice of IR cutoff (Hubble radius, event horizon, or conformal time), the FHDE model has been shown to produce quintessence-like and phantom-like behavior, admitting richer cosmological dynamics than the standard model. FHDE has been reconstructed via multiple scalar and gauge field models, yielding stable late-time acceleration while avoiding phantom instabilities \cite{Bidlan2025}. The authors investigated an LRS Bianchi type-I model with FHDE, constrained using MCMC analyses with Hubble, BAO, Pantheon, and SH0ES datasets, and found late-time accelerated expansion with statefinder diagnostics converging toward the $\Lambda$CDM fixed point \cite{Rathore2026}.

Parallel to dark-energy models, modifications of gravity-particularly those based on torsion-have gained significant interest. Teleparallel Gravity, formulated by Einstein, describes gravity in terms of torsion instead of curvature, and its modern generalization $f(T)$ gravity replaces the torsion scalar $T$ with a general function $f(T)$ in the gravitational action \cite{Ferraro2007,Ferraro2008,Li2011,Setare2012}. Unlike $f(R)$ gravity, which leads to fourth-order field equations, $f(T)$ gravity produces second-order equations, making it mathematically simpler and advantageous for cosmological applications. Numerous studies have demonstrated that $f(T)$ gravity can naturally generate cosmic acceleration without invoking exotic DE components \cite{Bamba2012,Bohmer2012}. Observationally constrained bulk viscous $f(T)$ cosmology exhibits a deceleration to acceleration transition with quintessence-like behavior which is consistent with the standard cosmological scenario \cite{Singh2026}.

The combination of FHDE with $f(T)$ gravity is particularly appealing. The interplay between the fractional HDE density and torsional modifications offers a wider range of dynamical behavior in cosmic evolution. However, the phase-space structure of such a combined model remains largely unexplored. Dynamical system analysis is a powerful tool in cosmology because it allows the investigation of asymptotic behavior, stable attractors, and transitions between different cosmological epochs by reducing the field equations into an autonomous system and studying its critical points \cite{Wainwright1997,Bahamonde2023,Leon2009}.
An interacting LRS Bianchi type-I model in $f(T)=T+\zeta T^{2}$ gravity exhibits equilibrium points corresponding to both quintessence and phantom dark energy regimes \cite{shivangi}. Dynamical system analysis of interacting as well as non-interacting $f(T)=\alpha T+\beta$ models shows stable critical points and late-time accelerated expansion which is consistent with observations \cite{Sarkar2025}. A dynamical system formulation of modified $f(R,G,T)=\alpha R^{l}+\beta G^{m}+\gamma T^{n}$ gravity has been developed using dimensionless variables, showing a late-time DE-dominated Universe with accelerated expansion and quintessence behavior, consistent with $\Lambda$CDM when confronted with Hubble, BAO, and Pantheon data \cite{MeeteiSingh2025}. Dynamical system analysis has been widely used to explore stability and late-time cosmic behavior in modified gravity and dark energy models \cite{Copeland1998,Samaddar2025,Samaddar2025FTB,Rathore2024,Singh2020}.

The objective of this work is to perform a complete phase-space analysis of fractional holographic dark energy within $f(T)$ gravity. We derive the autonomous system corresponding to the fractional HDE density, identify all critical points, analyze their stability using linear perturbations, and discuss the resulting cosmological implications. In particular, we investigate whether the model admits physically acceptable late-time attractors capable of describing the observed accelerated Universe. This paper is organized as follows. In sec. II, the basic formalism of the model is presented, where subsec. II.A discusses the FHDE with the Hubble horizon cutoff, and subsec. II.B briefly reviews the $f(T)$ gravity framework. In sec. III, we construct the dynamical variables and derive the autonomous system governing the cosmological evolution, followed by a detailed stability analysis of the critical points. subsec. III.A is devoted to the dynamical system formulation of FHDE in a non-interacting $f(T)$ gravity model, while subsec. III.B analyzes the existence and stability properties of the critical points. Finally, sec. IV summarizes the main results and presents the conclusions of this work.

\section{Basic formalism}

In this section, we present the theoretical background necessary for developing the phase-space analysis of FHDE within $f(T)$ gravity. We first introduce the formulation of FHDE, including the motivation for fractional generalizations and the dependence of the dark-energy density on the choice of infrared (IR) cutoff. This combined formalism provides the foundation for constructing the autonomous dynamical system explored in subsequent sections. We then summarize the essential features of teleparallel gravity and its modified extension $f(T)$ by highlighting the role of the $T$ and the resulting cosmological field equations.

\subsection{FHDE with Hubble cutoff}

The holographic principle asserts that the number of degrees of freedom of a physical system bounded by a surface is proportional to its area rather than its volume \cite{tHooft1993,Susskind1995}. This idea forms the theoretical basis of holographic dark energy (HDE), where the dark energy density is related to an infrared (IR) cutoff scale $L$ \cite{Li2004}. In the standard formulation, the total vacuum energy contained in a region of size $L$ should not exceed the mass of a black hole of the same size, leading to the bound $L^{3}\rho_{\mathrm{DE}}\leq LM_{p}^{2}$ \cite{Cohen1999}. Saturating this inequality gives the well-known expression for the HDE density
\begin{equation}
\rho_{\mathrm{HDE}}=3c^{2}M_{p}^{2}L^{-2},
\end{equation}
where $c$ is a dimensionless parameter and $M_{p}$ is the reduced Planck mass \cite{Li2004}. Although successful in describing dark energy phenomenology, this conventional framework does not incorporate possible nonlocal quantum gravitational corrections.

The incorporation of fractional calculus into cosmology provides a physically motivated extension of the standard theoretical framework, enabling one to account for nonlocality and memory effects that are expected to arise in gravitational and quantum systems \cite{Herrmann2011}. Fractional derivatives naturally encode such features and have found wide applications in different branches of physics. In particular, the fractal nature of quantum trajectories, emphasized in the path integral formulation of quantum mechanics by Feynman and Hibbs, suggests that fractional operators can effectively describe the underlying microscopic dynamics \cite{Feynman2010}. Furthermore, recent studies have shown that the fractional Wheeler--DeWitt equation modifies the thermodynamic properties of black holes and leads to a generalized entropy--area relation, thereby providing a strong theoretical motivation for introducing fractional corrections into cosmology \cite{Jalalzadeh2021}.

Fractional calculus extends classical differentiation and integration to non-integer orders and includes several well-known definitions such as the Liouville, Riemann, Caputo, and Riesz fractional derivatives, each appropriate for different physical contexts \cite{Herrmann2011,Ortigueira2017,Duarte2023}. These operators play a crucial role in describing self-similar and non-differentiable trajectories with fractal dimensions, and they naturally appear in space-fractional quantum mechanics formulated through path integrals over L\'evy-type trajectories \cite{Laskin2000}. In the limit $\alpha=2$, the usual Gaussian Brownian motion is recovered, while for $1<\alpha<2$ the dynamics reflects intrinsic fractal and nonlocal behavior.

When these fractional concepts are applied to gravity and cosmology, one arrives at the framework of fractional cosmology, where either the spacetime geometry or the dynamical field equations are generalized to fractional order \cite{Gonzalez2023,Socorro2023}. In particular, the fractional Wheeler--DeWitt equation yields a modified entropy--area relation of the form
\begin{equation}
S_{h}=C\,A^{\frac{\alpha+2}{2\alpha}}, \qquad 1<\alpha\leq 2,
\end{equation}
where $A$ is the horizon area and $\alpha$ denotes the fractional parameter \cite{Jalalzadeh2021}. The standard Bekenstein--Hawking entropy is recovered for $\alpha=2$, whereas for $1<\alpha<2$ fractional corrections modify the thermodynamic description of spacetime horizons.

Since the holographic bound connecting the ultraviolet (UV) cutoff and the IR cutoff explicitly depends on the entropy--area relation, this fractional modification alters the UV--IR correspondence and consequently changes the form of the dark energy density. Implementing the generalized entropy scaling into the holographic constraint leads to the FHDE density
\begin{equation}
\rho_{\mathrm{FHDE}}=3c^{2}M_{p}^{2}L^{-\frac{3\alpha-2}{\alpha}},
\end{equation}
which smoothly reduces to the standard HDE expression when $\alpha=2$. For $1<\alpha<2$, the scaling deviates from the usual inverse-square law, introducing fractional corrections that can significantly influence the late-time cosmological evolution.

In the present work, we adopt the Hubble radius as the IR cutoff, $L=H^{-1}$. With this choice, the FHDE density becomes
\begin{equation}
\rho_{\mathrm{FHDE}}=3c^{2}M_{p}^{2}H^{\frac{3\alpha-2}{\alpha}},
\end{equation}
demonstrating that the dark energy density now depends on a fractional power of the Hubble parameter determined by $\alpha$. The standard behavior $\rho\propto H^{2}$ is recovered for $\alpha=2$, while for $1<\alpha<2$ the modified exponent leads to a different dynamical evolution of dark energy. Therefore, the FHDE model with Hubble cutoff provides a consistent and theoretically motivated generalization of holographic dark energy, capable of incorporating nonlocal quantum gravitational effects and producing richer late-time cosmological dynamics.

The corresponding density parameter becomes
\begin{equation}
\Omega_{FHDE}
= \frac{\rho_{FHDE}}{3H^2}
= c^2\, M_p^2
H^{\frac{\alpha - 2}{\alpha}}.
\end{equation}

Assuming separate conservation of the FHDE component,
\begin{equation}
\dot{\rho}_{FHDE}
+ 3H(1+w_{FHDE})\rho_{FHDE} = 0,
\end{equation}
and using $\rho_{FHDE}\propto H^{(3\alpha-2)/\alpha}$,
we obtain
\begin{equation}
\dot{\rho}_{FHDE}
= \frac{3\alpha - 2}{\alpha}
H^{\frac{3\alpha - 2}{\alpha}-1}\dot{H},
\end{equation}
which leads to the equation-of-state parameter
\begin{equation}
w_{FHDE}
= -1
- \frac{3\alpha - 2}{3\alpha}
\frac{\dot{H}}{H^2}.
\label{wFHDE}
\end{equation}

Equation~(\ref{wFHDE}) explicitly shows that the dynamics of FHDE
are governed by the fractional parameter $\alpha$.
For $\alpha=2$ the standard HDE behavior is recovered,
whereas $1<\alpha<2$ modifies the effective equation of state
and enriches the late-time cosmological dynamics.

\subsection{ $f(T)$ gravity framework}
Teleparallel gravity is an alternative formulation of gravity in which torsion, rather than curvature, is used to describe the gravitational interaction. Instead of the Levi-Civita connection of General Relativity (GR), teleparallel gravity employs the Weitzenb\"{o}ck connection, which is curvature-free but possesses torsion. The fundamental dynamical variables are the tetrad (vierbein) fields $e^{A}{}_{\mu}$, which relate the spacetime metric $g_{\mu\nu}$ to the Minkowski metric $\eta_{AB}$ through
\begin{equation}
g_{\mu\nu} = e^{A}{}_{\mu} e^{B}{}_{\nu} \eta_{AB}.
\end{equation}

The torsion tensor corresponding to the Weitzenböck connection takes the form
\begin{equation}
T^{\tau}{}_{\mu\nu} = e_{A}{}^{\tau} \left( \partial_{\mu} e^{A}{}_{\nu} - \partial_{\nu} e^{A}{}_{\mu} \right),
\end{equation}
and from this, the torsion scalar $T$ is constructed as
\begin{equation}
T = S_{\tau}{}^{\mu\nu} T^{\tau}{}_{\mu\nu},
\end{equation}
where $S_{\tau}{}^{\mu\nu}$ is the superpotential tensor defined by
\begin{equation}
S_{\tau}{}^{\mu\nu} = \frac{1}{2}\left(K^{\mu\nu}{}_{\tau} + \delta^{\mu}_{\tau}\, T^{\gamma\nu}{}_{\gamma} - \delta^{\nu}_{\tau}\, T^{\gamma\mu}{}_{\gamma}\right),
\end{equation}
and $K^{\mu\nu}{}_{\tau}$ is the contortion tensor,
\begin{equation}
K^{\mu\nu}{}_{\tau} = -\frac{1}{2}\left(T^{\mu\nu}{}_{\tau} - T^{\nu\mu}{}_{\tau} - T_{\tau}{}^{\mu\nu} \right).
\end{equation}

The teleparallel equivalent of General Relativity (TEGR) is obtained from the action
\begin{equation}
S_{\text{TEGR}} = \frac{1}{2} \int d^4x\, e\, T + \int d^4x\, e\, \mathcal{L}_m,
\end{equation}
where $e = \det(e^A{}_{\mu}) = \sqrt{-g}$ is the determinant of the tetrad, and $\mathcal{L}_m$ is the matter Lagrangian density.  

In the modified teleparallel theory known as $f(T)$ gravity, the torsion scalar $T$ is generalized to an arbitrary function $T+f(T)$:
\begin{equation}
S = \frac{1}{2} \int d^4x\, e\, [T+f(T)] + \int d^4x\, e\, \mathcal{L}_m.
\end{equation}

By performing a variation of the action with respect to the tetrad, one obtains the gravitational field equations:
\begin{align}
&e^{-1}\partial_\mu \left( e\, e^\tau_{\ A} S_\tau^{\ \mu\nu} \right)\left[ 1 + f_T \right]
- e^\lambda_{\ A} T^\tau_{\ \mu\lambda} S_\tau^{\ \nu\mu} \left[ 1 + f_T \right] 
+ e^\tau_{\ A} S_\tau^{\ \mu\nu} \, \partial_\mu(T) f_{TT}
- \frac{1}{4} e^\nu_{\ A} \left[ T + f(T) \right]
= k\, e^\tau_{\ A} \mathcal{T}_\tau^{\ \nu}
\end{align}

where $f_T \equiv \partial f/\partial T$ and $f_{TT} \equiv \partial^2 f/\partial T^2$, and $\mathcal{T}_{\tau}{}^{\nu}$ is the matter energy-momentum tensor. For a spatially flat Friedmann-Robertson-Walker (FRW) universe with scale factor $a(t)$, the tetrad field can be chosen as $e^{A}{}_{\mu} = \text{diag}(1,a,a,a)$, yielding the torsion scalar
\begin{equation}
T = -6H^2,
\end{equation}
where $H = \dot{a}/a$ is the Hubble parameter. The modified Friedmann equations in $f(T)$ gravity can be written as
\begin{equation}
3H^2 = \rho + \frac{1}{2}\left( 2T f_T - f(T) \right),
\label{FE1}
\end{equation}

\begin{equation}
-2\dot{H} = \frac{\rho + p}{1 + f_T + 2T f_{TT}},
\label{FE2}
\end{equation}
where $f_T \equiv \frac{df}{dT}$ and $f_{TT} \equiv \frac{d^2 f}{dT^2}$.

These equations \ref{FE1} and \ref{FE2} can be recast into a form analogous to the standard
Friedmann equations of General Relativity by introducing effective
energy density and pressure terms defined as
\begin{equation}
3H^2 = \rho_{\rm eff},
\end{equation}

\begin{equation}
-2\dot{H} = \rho_{\rm eff} + p_{\rm eff}.
\end{equation}
where effective energy density $(\rho_{\rm eff})$ and pressure $(p_{\rm eff})$ are given by
\begin{align}
\rho_{\rm eff} &= \rho + \frac{1}{2}\left( 2T f_T - f(T) \right), \\
\rho_{\rm eff} +p_{\rm eff} &= \frac{\rho + p}{1 + f_T + 2T f_{TT}}.
\end{align}

The additional contribution arising from torsion can be interpreted
as an effective dark energy component. Accordingly, the torsional
dark energy density is defined as
\begin{equation}
\rho_T = \frac{1}{2}\left( 2T f_T - f(T) \right).
\end{equation}

Thus, $f(T)$ gravity provides an effective dark-energy component through torsion and leads to modified cosmological dynamics that can be explored via dynamical system techniques. The effective EoS parameter and the deceleration parameter are employed to describe the cosmological dynamics and are defined by
\begin{align}
\omega_{\rm{eff}} &=\frac{p_{\rm eff}}{\rho_{\rm eff}}= -1 - \frac{2}{3}\frac{\dot{H}}{H^{2}}\label{w},\\
\text{and} \,\,\,\,\,\,\,\,\,\,\, q &= -1 - \frac{\dot{H}}{H^{2}}\label{q}.
\end{align}

\section{Dynamical variables, autonomous system and stability of critical points}

In order to study the qualitative evolution of cosmological models, it is often useful to rewrite the field equations in the form of an closed dynamical system in $\mathbb{R}^n$. A dynamical system is defined as a set of first-order differential equations of the form
\begin{equation}
\mathbf{x}' = \mathbf{f}(\mathbf{x}),
\end{equation}
where $\mathbf{x} \in \mathbb{R}^n$, $\mathbf{f} : \mathbb{R}^n \rightarrow \mathbb{R}^n$ is a continuously differentiable vector-valued function, and derivatives marked by a prime are evaluated w.r.t. the logarithmic scale factor $N \equiv \ln a$. The solutions of such a system describe trajectories in phase space, with each point representing a specific cosmological state. A critical (or equilibrium) point $\mathbf{x}_c$ is defined by
\begin{equation}
\mathbf{f}(\mathbf{x}_c) = 0,
\end{equation}
which corresponds to a cosmological regime where the dynamical variables remain constant.

The stability of a critical point is determined by linearizing the system around $\mathbf{x}_c$. Let $\mathbf{u} = \mathbf{x} - \mathbf{x}_c$ represent a perturbation. Expanding $\mathbf{f}(\mathbf{x})$ around $\mathbf{x}_c$ and retaining only first-order terms yields the linearized system
\begin{equation}
\mathbf{u}' = J(\mathbf{x}_c)\, \mathbf{u},
\end{equation}
where $J(\mathbf{x}_c)$ is the Jacobian matrix of $\mathbf{f}$ evaluated at the critical point,
\begin{equation}
J_{ij}(\mathbf{x}_c) = \left. \frac{\partial f_i}{\partial x_j} \right|_{\mathbf{x}_c}.
\end{equation}
The eigenvalues of $J(\mathbf{x}_c)$ determine the nature of the critical point. Stability is ensured if the real parts of all eigenvalues are negative; conversely, the critical point becomes unstable when any eigenvalue exhibits a positive real component. Mixed signs of real parts of all eigenvalues indicate a saddle point. These results follow from the Hartman–Grobman theorem, which guarantees that the nonlinear system behaves locally like its linearization near hyperbolic critical points \cite{Strogatz1994,Perko2001}.

The dynamical systems approach has been widely used in cosmology to classify different evolutionary regimes dominated by radiation, matter, or dark energy \cite{Wainwright1997,Leon2009}. By converting the cosmological equations into an autonomous system in suitable dimensionless variables, one can obtain a comprehensive view of the asymptotic behavior and global structure of the cosmological dynamics. In what follows, we apply this formalism to fractional holographic dark energy in $f(T)$ gravity and construct the corresponding autonomous system.

\subsection{Dynamical system formulation of FHDE in a non-interacting $f(T)$ gravity model}

 In this part of the analysis, we adopt a quadratic modification of teleparallel gravity defined via the function
\begin{equation}
f(T) = \beta T^{2},
\end{equation}
where $\beta$ is a free parameter of the model. This particular choice represents the simplest nonlinear extension beyond the standard TEGR formulation, introducing higher-order torsion contributions into the gravitational action. The parameter $\beta$ governs the magnitude of these deviations and may significantly affect the cosmic dynamics, especially during the late-time acceleration phase. Substitution of this form into the modified Friedmann equations enables us to construct the corresponding autonomous dynamical system for further analysis.

To rewrite the cosmological evolution equations within an autonomous phase-space formulation, we introduce the following set of dimensionless variables:

\begin{equation}
X=\frac{\rho_m}{3H^2},\, Y=\frac{\rho_r}{3H^2}, \, Z=\frac{\rho_{FHDE}}{3H^2} \, \, \text{and} \, \, \zeta=\frac{\rho_T}{3H^2}.
\end{equation}

where the effective energy density of the Universe is expressed as the combined contributions from matter, radiation, the FHDE and the torsional DE component:

\begin{equation}
\rho_{\rm eff}=\rho +\rho_{T}=\rho_m+\rho_r+\rho_{FHDE}+\rho_{T}.
\end{equation}

In this framework, $\rho_m$ denotes the energy density of non\text{-}relativistic matter, while $\rho_r$ corresponds to the energy density of the radiation component. Their associated pressures follow the standard relations: (i) the pressure of matter is $p_m = 0$, reflecting its negligible relativistic effects; (ii) the pressure of radiation is given by $p_r = \rho_r/3$, in accordance with the relativistic EoS; (iii) for the FHDE sector, the pressure is defined as $p_{FHDE} = \omega_{FHDE}\rho_{FHDE}$, where $\omega_{FHDE}$ represents the EoS parameter of the FHDE and (iv) for the torsional DE sector, the pressure is defined as $p_{T} = \omega_{T}\rho_{T}$, where $\omega_{T}$ represents the EoS parameter of the torsional DE. The corresponding energy conservation equations for each component take the following form:
\begin{align}
\dot{\rho}_{m} + 3H\rho_{m} &= 0, \label{cons_m} \\
\dot{\rho}_{r} + 4H\rho_{r} &= 0, \label{cons_r} \\
\dot{\rho}_{FHDE} + 3H(1+\omega_{FHDE})\rho_{FHDE} &= 0. \label{cons_f}\\
\dot{\rho}_{T} + 3H(1+\omega_{T})\rho_{T} &= 0. \label{cons_T}
\end{align}

Density parameters are given as:
\begin{equation}
\Omega_m=\frac{\rho_m}{3H^2},\, \Omega_r=\frac{\rho_r}{3H^2}, \, \Omega_{FHDE}=\frac{\rho_{FHDE}}{3H^2} \, \text{and} \, \Omega_T=\frac{2Tf_T - f(T)}{3H^2}=18\beta H^2.
\end{equation}

Note that $\Omega_T$ is the density parameter for the effective dark-energy component through torsion. Now, using eqs. (\ref{FE1}) and (\ref{FE2}), we have

\begin{equation}
\zeta=1-X-Y-Z.
\end{equation}

\begin{equation}
\frac{\dot{H}}{H^2}=-\frac{1}{2}\left[\frac{3X+4Y+3(1+\omega_{FHDE})Z}{1-2\zeta}\right].
\end{equation}

Substituting the expression for $\omega_{FHDE}$ from equation (\ref{wFHDE}) into the above equation and simplifying the resulting relation allows us to derive an explicit expression for the dimensionless quantity $\dot{H}/H^{2}$, which is given by:

\begin{equation}
\frac{\dot{H}}{H^2}=\frac{3X+4Y}{4\zeta-2+\left(\frac{3\alpha-2}{3\alpha}\right)Z} . \label{HH}
\end{equation}

Using this relation, we can express both the deceleration parameter and the effective EoS parameter given in equations \ref{w} and \ref{q} in closed analytical form as follows:
\begin{align}
\omega_{eff} & = -1 - \frac{2}{3}\frac{\dot{H}}{H^{2}}=\frac{-6+6X+4Y+\left(\frac{9\alpha+2}{\alpha}\right)Z}{6-12X-12Y-\left(\frac{9\alpha+2}{\alpha}\right)Z}\\
q&= -1 - \frac{\dot{H}}{H^{2}}=\frac{-2+X+\left(\frac{9\alpha+2}{3\alpha}\right)Z}{2-4X-4Y-\left(\frac{9\alpha+2}{3\alpha}\right)Z} \\
\end{align} 

Also, from eqn. (\ref{wFHDE}) EoS for FHDE becomes  
\begin{equation}
w_{FHDE} = -1 + \frac{3\alpha - 2}{3\alpha}\left[\frac{3X+4Y}{2-4X-4Y-\left(\frac{9\alpha+2}{3\alpha}\right)Z}\right].
\end{equation}

Since $\zeta$ depends on the other dynamical quantities of the system, we construct the autonomous system using the variables $X$, $Y$, and $Z$. The resulting dynamical system can therefore be written as follows:
\begin{align}
X' &= -3X-2\left[\frac{3X+4Y}{2-4X-4Y-\left(\frac{9\alpha+2}{3\alpha}\right)Z}\right]X\\
Y' &=-4Y-2\left[\frac{3X+4Y}{2-4X-4Y-\left(\frac{9\alpha+2}{3\alpha}\right)Z}\right]Y\\
Z' &=\left(\frac{\alpha-2}{\alpha}\right)\left[\frac{3X+4Y}{2-4X-4Y-\left(\frac{9\alpha+2}{3\alpha}\right)Z}\right]Z
\end{align}

\subsection{Critical points and their stabilities}

To determine critical points (CP), we solve the autonomous system by equating $X'=0,Y'=0 \,\text{and}\, Z'=0$. The resulting CP together with their associated cosmic parameters are summerized in Table (\ref{tab}).  We obtain the eigenvalues by evaluating the Jacobian matrix at each CP. The corresponding eigenvalues and stability conditions are presented in Table (\ref{tab2}).

\begin{figure}[h!]
    \centering
    \begin{subfigure}{0.4\textwidth}
        \centering
        \includegraphics[width=\linewidth]{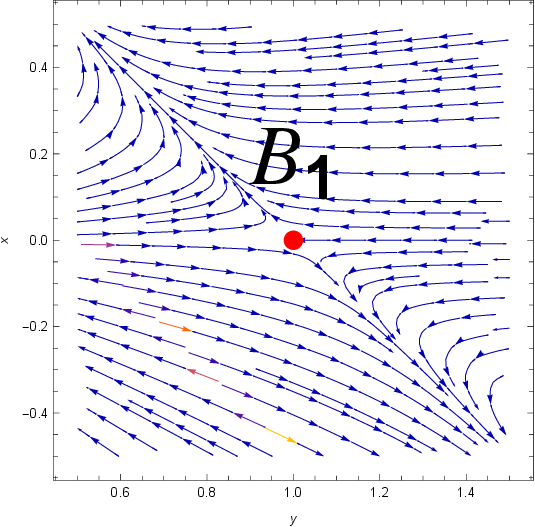}
        \caption{}\label{b1}
    \end{subfigure}

     \medskip
     
    \begin{subfigure}{0.4\textwidth}
        \centering
        \includegraphics[width=\linewidth]{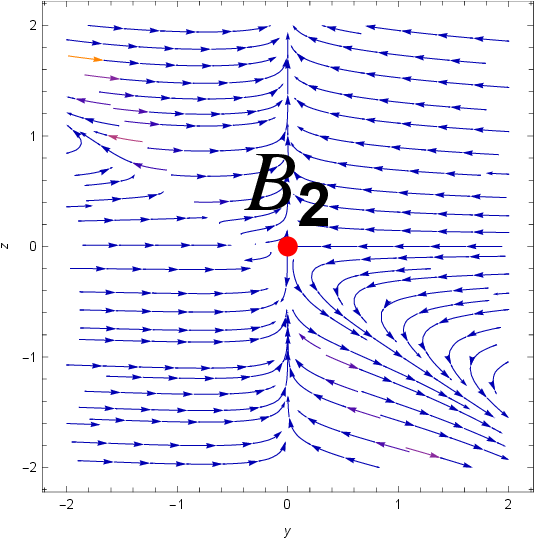}
        \caption{}\label{b2}
    \end{subfigure}
\hfill
\begin{subfigure}{0.4\textwidth}
        \centering
        \includegraphics[width=\linewidth]{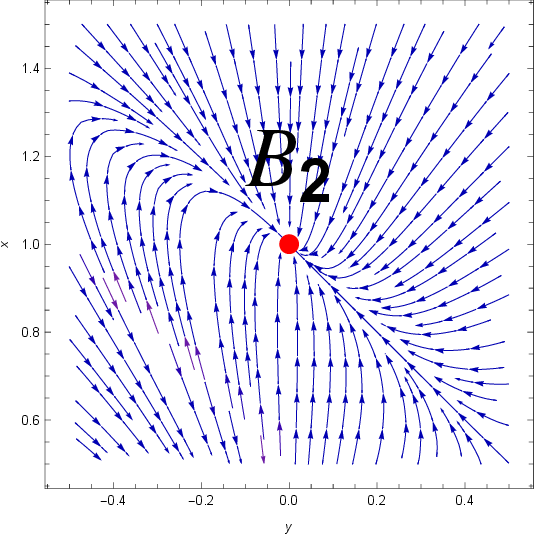}
        \caption{}\label{b22}
    \end{subfigure}
    
    \medskip

    \begin{subfigure}{0.4\textwidth}
        \centering
        \includegraphics[width=\linewidth]{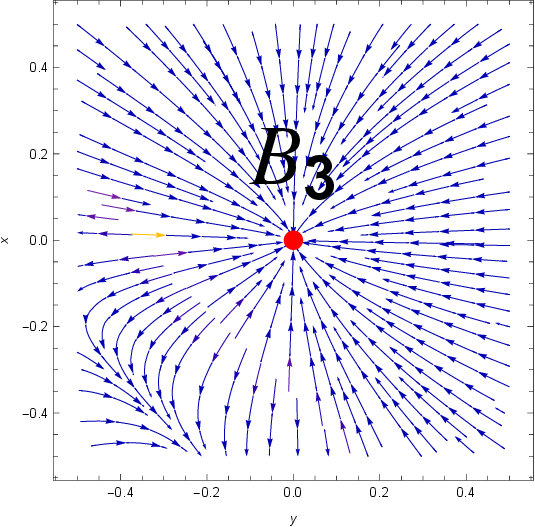}
        \caption{}\label{b3}
    \end{subfigure}
    \hfill
    \begin{subfigure}{0.4\textwidth}
        \centering
        \includegraphics[width=\linewidth]{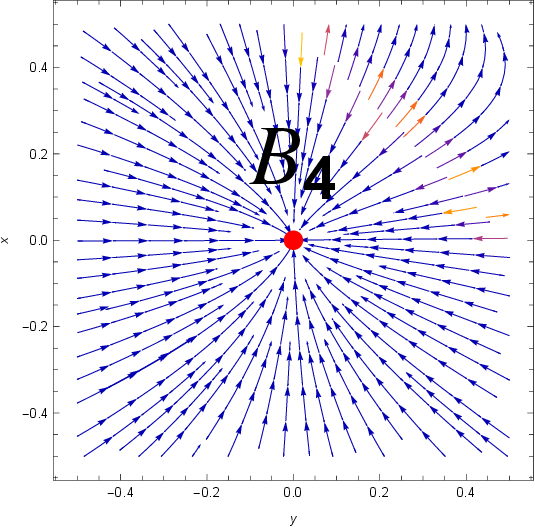}
        \caption{}\label{b4}
    \end{subfigure}

    \caption{Phase portrait diagrams for the autonomous dynamical system with $\alpha=1.5$}
\end{figure}

\underline{\bf CP $B_1$:} This critical point always exists. As shown in Table (\ref{tab2}), the coexistence of both positive and negative eigenvalues confirms that it is a saddle point. At this critical point, the density parameters satisfy $\Omega_m = 0$, $\Omega_r = 1$, $\Omega_{\mathrm{FHDE}} = 0$, and $\Omega_T = 0$, indicating that the cosmic dynamics are entirely dominated by radiation, while the contributions from matter, fractional holographic dark energy, and torsion-induced dark energy are negligible. The effective EoS parameter takes the value $\omega_{\mathrm{eff}} = \frac{1}{3}$, and the deceleration parameter is $q = 1$, confirming the standard radiation-dominated expansion. Since $\Omega_{\mathrm{FHDE}} = 0$, the FHDE component does not contribute to the effective equation of state and therefore has no influence on the cosmic dynamics. Consequently, the specific value of the FHDE EoS parameter is dynamically irrelevant at this critical point. The corresponding phase portrait for this critical point is presented in Fig.~\ref{b1} with $\alpha=1.5$. In this case, the scale factor follows the power-law behavior $a(t) \propto (2t + k)^{1/2}$, where $k$ is a constant of integration. Accordingly, this CP represents a cosmological phase with decelerated expansion which is the characteristic of a radiation-dominated regime. For $\alpha=2$, the eigenvalue $\frac{2(2-\alpha)}{\alpha}$ vanishes, indicating that the corresponding critical point becomes non-hyperbolic. Nevertheless, since the remaining eigenvalues are $1$ and $-4$, the point retains a saddle character in the cosmological phase space. 

\begin{table}[h!]
\centering
\begin{tabular}{|c|c|c|c|c|c|c|c|c|}
\hline
\hline
 & $X=\Omega_m$ & $Y=\Omega_r$ & $Z=\Omega_{FHDE}$ & $\zeta=\Omega_T$ & $q$ & $\omega_{\mathrm{eff}}$ & $\omega_{FHDE}$ & Existence \\ \hline
 \hline
 \hline
$B_1$    &   $0$      &     $1$     &     $0$     &     $0$     &    $1$  &  $\frac{1}{3}$  &  \tablefootnote{At this critical point, the FHDE density parameter vanishes, 
i.e., $\Omega_{\mathrm{FHDE}} = 0$. Consequently, the FHDE component is dynamically absent and does not contribute to the effective EoS or the cosmic evolution. Therefore, the explicit value of the FHDE EoS parameter $\omega_{\mathrm{FHDE}}$ is not physically relevant and is not evaluated at this point.} &Always\\ \hline
$B_2$    &    $1$      &    $0$      &     $0$     &      $0$    &    $\frac{1}{2}$  &  $0$  &   \tablefootnote{Same as above.} & Always\\ \hline
$B_3$    &    $0$       &    $0$      &     $1$     &     $0$     &    $-1$  &  $-1$  &   $-1$ &Always\\ \hline
$B_4$    &    $0$       &      $0$    &     $0$     &     $1$     &   $-1$   &  $-1$  &  \tablefootnote{Same as above.}  & Always\\ \hline
\end{tabular}
\caption{CP and their associated cosmic parameters.}
\label{tab}
\end{table}

\begin{table}[h!]
\centering
\begin{tabular}{|c|c|c|c|c|}
\hline
\hline
CP & $\lambda_1$ & $\lambda_2$ & $\lambda_3$ &  Stability \\ \hline
\hline
\hline
$B_1$     &   $\frac{2(2-\alpha)}{\alpha}$       &       $1$      &        $-4$    &         Saddle; $1<\alpha\le2$      \\ \hline
$B_2$     &     $\frac{3(2-\alpha)}{2\alpha}$     &       $-3$      &        $-1$   &      Saddle; $1<\alpha<2$ \& Stable: $\alpha=2$            \\ \hline
$B_3$     &     $0$      &     $-4$      &      $-3$     &          Stable;   $1<\alpha\le2$        \\ \hline
$B_4$     &       $-4$    &     $-3$      &      $0$   &           Stable;    $1<\alpha\le2$       \\ \hline
\end{tabular}
\caption{Eigenvalues and stability conditions for the CP.}
\label{tab2}
\end{table}

\begin{table}[h!]
\centering
\begin{tabular}{|c|c|c|}
\hline
CP & evolution of $a(t)$ & Dominant Component \\ 
\hline
 $B_1$ & $a(t)\propto (2t+k)^{1/2}$ & Radiation \\ 
\hline
 $B_2$ &  $a(t)\propto (\frac{3}{2}t+k)^{2/3}$ & Matter \\ 
\hline
 $B_3$ & $a(t)\propto e^{kt}$ & FHDE \\ 
\hline
 $B_4$ & $a(t)\propto e^{kt}$ &  torsion-induced DE\\ 
\hline
\end{tabular}
\caption{Evaluation of $a(t)$ at different CP, including the dominant energy component}
\label{}
\end{table}

\underline{\bf CP $B_2$:} This critical point always exists. As shown in Table (\ref{tab2}), the eigenvalue of $\frac{3(2-\alpha)}{2\alpha}$ is positive for $1<\alpha<2$. The coexistence of both positive and negative eigenvalues confirms that it is a saddle point. At this CP, the density parameters satisfy $\Omega_m = 1$, $\Omega_r = 0$, $\Omega_{\mathrm{FHDE}} = 0$, and $\Omega_T = 0$, indicating that the cosmic dynamics are entirely dominated by matter, while the contributions from radiation, fractional holographic dark energy, and torsion-induced dark energy are negligible. The effective EoS parameter takes the value $\omega_{\mathrm{eff}} = 0$, and the deceleration parameter is $q = \frac{1}{2}$, confirming the standard radiation-dominated expansion. Since $\Omega_{\mathrm{FHDE}} = 0$, the FHDE component does not contribute to the effective equation of state and therefore has no influence on the cosmic dynamics. Consequently, the specific value of the FHDE EoS parameter is dynamically irrelevant at this critical point. The corresponding phase portrait for this critical point is presented in Fig.~\ref{b2} with $\alpha=1.5$. For $\alpha=2$, the eigenvalue of $\frac{3(2-\alpha)}{2\alpha}$ is $0$. The occurrence of a zero eigenvalue indicates that this critical point is non-hyperbolic, meaning that linear stability analysis alone is insufficient to determine its dynamical character. From the phase portrait shown in Fig.~\ref{b22}, it is evident that all trajectories asymptotically converge toward the critical point, confirming that it represents a stable CP. In this case, the $a(t)$ evolves according to the power-law form $a(t) \propto \left(\frac{3}{2}t + k\right)^{2/3}$, where $k$ is a constant of integration. This CP therefore corresponds to a cosmological phase with decelerated expansion, characteristic of a matter-dominated regime. 

\underline{\bf CP $B_3$:} This critical point always exists. As presented in Table (\ref{tab2}), the occurrence of a zero eigenvalue indicates that this critical point is non-hyperbolic, meaning that linear stability analysis alone is insufficient to determine its dynamical character. From the phase portrait shown in Fig.~\ref{b3}, it is evident that all trajectories asymptotically converge toward the critical point, confirming that it represents a late-time stable attractor of the dynamical system. At this point, the density parameters satisfy $\Omega_m = 0$, $\Omega_r = 0$, $\Omega_{\mathrm{FHDE}} = 1$, and $\Omega_T = 0$. Consequently, the late-time cosmic evolution is completely dominated by FHDE, while the contributions from matter, radiation, and torsion-induced DE become negligible. Moreover, both the effective EoS parameter and the FHDE EoS take the values $\omega_{\mathrm{eff}} = -1$ and $\omega_{\mathrm{FHDE}} = -1$, respectively, while the deceleration parameter is $q = -1$. These values indicate that the CP denotes to a de Sitter phase of the Universe. This result demonstrates that the model naturally evolves toward a FHDE-dominated accelerated expansion, independent of the initial conditions. Furthermore, the $a(t)$ exhibits an exponential behavior of the form $a(t) \propto e^{k t}$, where $k$ is an integration constant. Hence, this CP represents a de Sitter–type cosmological solution, marked by a phase of accelerated expansion driven by an exponentially growing scale factor.

\underline{\bf CP $B_4$:} This critical point always exists. As presented in Table (\ref{tab2}), the occurrence of a zero eigenvalue indicates that this critical point is non-hyperbolic, meaning that linear stability analysis alone is insufficient to determine its dynamical character. From the phase portrait shown in Fig.~\ref{b4}, it is evident that all trajectories asymptotically converge toward the critical point, confirming that it represents a late-time stable attractor of the dynamical system. At this CP, the density parameters satisfy $\Omega_m = 0$, $\Omega_r = 0$, $\Omega_{\mathrm{FHDE}} = 0$, and $\Omega_T = 1$, where $\Omega_T$ denotes the energy density contribution arising from torsion. Consequently, the late-time cosmic evolution is completely dominated by torsional effects, while the contributions from matter, radiation, and FHDE become negligible. Furthermore, the effective EoS parameter and the deceleration parameter take the values $\omega_{\mathrm{eff}} = -1$ and $q = -1$, respectively, indicating that this CP corresponds to a de Sitter phase. Since $\Omega_{\mathrm{FHDE}} = 0$, the FHDE component does not contribute to the effective equation of state and therefore has no influence on the cosmic dynamics. Consequently, the specific value of the FHDE EoS parameter is dynamically irrelevant at this critical CP. Consequently, the late-time accelerated expansion is entirely driven by the torsional sector, demonstrating that the model naturally evolves toward a torsion-dominated de Sitter attractor, independent of the initial conditions. This result demonstrates that the model naturally evolves toward a torsion-dominated accelerated expansion, independent of the initial conditions. Furthermore, the scale factor is found to follow the exponential law $a(t) \propto e^{k t}$, where $k$ is a constant of integration. Consequently, this CP corresponds to a de Sitter--type cosmological solution, characterized by an accelerated expansion with an exponentially evolving scale factor.

Overall, the phase-space analysis confirms that the model can successfully account for the standard thermal history of the Universe, encompassing the radiation-dominated phase, the matter-dominated phase, and the subsequent DE-dominated accelerated expansion, which can be schematically represented as
\begin{equation}
\text{Radiation-dominated phase} \;\longrightarrow\; 
\text{Matter-dominated phase} \;\longrightarrow\; 
\text{DE-dominated accelerated phase}.
\end{equation}

\section{Conclusion}
In this work, we investigate the cosmological evolution of the FHDE model within the framework of $f(T)=\beta T^2$ gravity in a spatially flat FLRW metric. By introducing suitable dimensionless variables, the cosmological field equations are rewritten as an autonomous system, allowing a systematic analysis of the phase-space structure and the stability of the CP.

The critical point $B_1$ corresponds to a radiation-dominated saddle point of the dynamical system. At this point, the cosmic dynamics is entirely governed by radiation, with the contributions from matter, FHDE, and torsion-induced DE becoming negligible. The values $\omega_{\mathrm{eff}}=\frac{1}{3}$ and $q=1$ confirm that this critical point represents a decelerated phase of the Universe and plays the role of a transient early-time epoch.

The critical point $B_2$ corresponds to a matter-dominated saddle point of the dynamical system. At this point, the cosmic dynamics is entirely governed by matter, with the contributions from radiation, FHDE and torsion-induced DE becoming negligible. The values $\omega_{\mathrm{eff}}=0$ and $q=\frac{1}{2}$ confirm that this critical point represents a decelerated phase of the Universe.

The critical point $B_3$ corresponds to a late-time de Sitter attractor of the dynamical system. Although the presence of a zero eigenvalue makes this point non-hyperbolic, the phase-space analysis shows that all trajectories asymptotically approach it, implying late-time stability. At this point, the cosmic dynamics is entirely governed by FHDE, with the contributions from matter, radiation, and torsion-induced DE becoming negligible. The values $\omega_{\mathrm{eff}}=\omega_{\mathrm{FHDE}}=-1$ and $q=-1$ confirm that this critical point represents an accelerated de Sitter phase of the Universe.

The critical point $B_4$ corresponds to a late-time de Sitter attractor of the dynamical system. Although the presence of a zero eigenvalue makes this point non-hyperbolic, the phase-space analysis shows that all trajectories asymptotically approach it, implying late-time stability. At this point, the cosmic dynamics is entirely governed by torsion-induced DE, with the contributions from matter, radiation, and FHDE becoming negligible. The values $\omega_{\mathrm{eff}}=-1$ and $q=-1$ confirm that this critical point represents an accelerated de Sitter phase of the Universe.

Overall, the dynamical analysis indicates that the $f(T)$ model can reproduce the standard sequence of cosmological epochs, including radiation dominated, matter dominated, and a late-time DE–dominated phase. The corresponding critical points are associated with distinct stages of cosmic evolution, such as decelerated radiation and matter eras and a late-time accelerated de Sitter state. These results suggest that the proposed model provides a consistent theoretical framework capable of describing the past, present, and future evolution of the Universe within a unified dynamical approach.

 \end{document}